# Resolution of the Abraham-Minkowski Controversy

Masud Mansuripur

College of Optical Sciences, The University of Arizona, Tucson, Arizona 85721

**Abstract**. The momentum of light inside ponderable media has an electromagnetic part and a mechanical part. The local and instantaneous density of the electromagnetic part of the momentum is given by the Poynting vector divided by the square of the speed of light in vacuum, irrespective of the nature of the electromagnetic fields or the local or global properties of the material media. The mechanical part of the momentum is associated with the action of the electromagnetic field on the atomic constituents of the media, as specified by the Lorentz law of force. Proper interpretation and application of the Maxwell-Lorentz equations within the material bodies as well as at their surfaces and interfaces is *all* that is needed to obtain a complete picture of the momentum of light, including detailed numerical values at each and every point in space and time. That the Abraham-Minkowski controversy surrounding the momentum of light inside material media has persisted for nearly a century is due perhaps to an insufficient appreciation for the completeness and consistency of the *macroscopic* Maxwell-Lorentz theory, inadequate treatment of the electromagnetic force and torque at the material boundaries, and an undue emphasis on the necessity of coupling the equations of electrodynamics to those of the theory of elasticity for proper treatment of mechanical momentum. The present paper reports the resolution of the Abraham-Minkowski controversy within the framework of the classical theory of electrodynamics, without resort to such complicating and ultimately unnecessary factors as pseudo-momentum, special surface forces, alternative energy-momentum tensors, and hidden momenta, that have caused so much confusion for such a long period of time.

**1. Introduction**. The Abraham-Minkowski controversy dates back to the early years of the 20$^{th}$ Century, when Hermann Minkowski's electromagnetic theory indicated that a light pulse of energy $\mathcal{E}_{pulse}$ within a dielectric medium of refractive index $n$ would have a momentum of $n\mathcal{E}_{pulse}/c$ ($c$ being the speed of light in vacuum) whereas Max Abraham's theory gave a value of $\mathcal{E}_{pulse}/(nc)$ for the same momentum [1,2]. Over the years, many theoretical arguments have been advanced [3-27] and experimental evidence provided [28-38] in support of one theory over the other, but the controversy persists. As it turns out, the Abraham-Minkowski conundrum can be fully resolved if the Maxwell-Lorentz equations of classical electrodynamics are properly interpreted and correctly applied. The solution to the problem of optical momentum in ponderable media is fully consistent with the conservation laws and does not require any new physics or complicated hypotheses. The solution applies to any type of material medium, whether or not the medium is homogeneous, transparent, dispersive, birefringent, magnetic, non-linear, etc.; in particular, it is applicable to negative-index media, where the dielectric permittivity $\varepsilon$ and magnetic permeability $\mu$ are both negative-valued.

It is not our intention here to describe in mathematical detail the most general solution to the problem of optical momentum in ponderable media. Rather, we shall describe the main features of the solution and give reference to the literature for rigorous derivations [18−21, 39−53]. We also provide qualitative descriptions of the mechanical momentum imparted to material media in their interactions with the electromagnetic field. (It is our opinion that a clear, step-by-step picture of the mechanism by which electromagnetic and mechanical momenta are exchanged is crucial to an understanding of the role played by mechanical momentum in resolving the Abraham-Minkowski controversy.) The main features of the momentum of light as revealed by a comprehensive analysis of the Maxwell-Lorentz equations without resort to unwarranted approximations or unsubstantiated hypotheses may be summarized as follows:

1) In the simple case of a light pulse propagating inside a transparent, dispersionless dielectric of refractive index $n$, the total momentum associated with the light pulse is $\frac{1}{2}(n+n^{-1})\mathcal{E}_{pulse}/c$. This momentum consists of two parts, an electromagnetic part that is given by the Abraham formula $p_{EM} = \mathcal{E}_{pulse}/(nc)$, and a mechanical part that is given by $p_{mech} = \frac{1}{2}(n-n^{-1})\mathcal{E}_{pulse}/c$.

2) In the most general case, one must solve Maxwell's macroscopic equations to determine the electromagnetic field distribution $\boldsymbol{E}(\boldsymbol{r},t)$ and $\boldsymbol{H}(\boldsymbol{r},t)$ throughout the region of interest. The electromagnetic momentum density is *always* given by $\boldsymbol{E} \times \boldsymbol{H}/c^2$, which is commonly associated with the name of Abraham. This expression is completely general and does not depend on whether the medium within which the electromagnetic momentum density is sought is homogeneous, absorptive, dispersive, isotropic, magnetic, linear, etc. The formula even applies to evanescent waves, such as those found in the cladding region of optical fibers or other dielectric waveguides [51,52].

3) The mechanical force exerted by light on material media is *always* found from the generalized Lorentz law of force; no *ad hoc* alternative expressions for the force density are necessary. In particular, the forces at surfaces and at interfaces between adjacent media must all be calculated in accordance with the Lorentz law, with the boundary conditions derived from Maxwell's macroscopic equations [39–43]. The Lorentz force fully specifies the exchange of mechanical momentum between the light and the material media. There is no need for invoking special surface forces [11,12], pseudo-momentum [15], hidden momentum [8,9], or any other artificial device to account for momentum conservation. (Pfeifer *et al*'s treatment of the Jones experiments [28,33] under the Abraham energy-momentum tensor in Sec. VIII A of [26] similarly avoids such artificial devices. This is indicative of the common ground underlying the two approaches to radiation pressure problems, one based on the Lorentz law of force, the other on the energy-momentum tensor formalism.)

4) The mechanical force exerted by electromagnetic waves on material media, as obtained by applying the generalized Lorentz law of force, is precisely equal to minus the time-rate-of-change of the electromagnetic momentum within the entire system, namely, $F^{(\text{total})} = -\mathrm{d}p_{\text{EM}}^{(\text{total})}/\mathrm{d}t$ [51,52]. It is this precise relationship between mechanical force and the time-rate-of-change of electromagnetic momentum that ensures the conservation of linear momentum at each and every instant of time throughout the system.

5) The same considerations as above apply to optical and mechanical angular momentum in conjunction with the mechanical torque exerted by a light pulse. Once again, the macroscopic Maxwell equations, the generalized Lorentz law, and the corresponding boundary conditions are all that one needs in order to calculate the electromagnetic angular momentum and the torque exerted by fields on material media [20,42,49].

In general, of course, one must solve Maxwell's electrodynamics equations in conjunction with the elastic equations that describe the mechanical response of material media to electromagnetic forces and torques. The change of pressure and density as well as material flow will, in turn, modify the electromagnetic properties of the media, which the solver of Maxwell's equations must take into account in a self-consistent way. What we present in this paper, however, pertains to the extreme case where the media of interest are sufficiently rigid and massive that elastic interactions between adjacent molecules, acoustic wave generation and propagation, and mechanical motion of the media may be safely ignored. When a stationary object of mass $M$ picks up a momentum $p$ as a result of interaction with a light pulse, its acquired kinetic energy will be $p \cdot p/(2M)$. An implicit assumption in the following discussions is that $M$ is large enough that ignoring the transfer of this kinetic energy, from the light pulse to the object, does *not* affect the main conclusions. Similarly, we do not require infinite rigidity from the material media; only that their elastic properties be such that the energies involved in the generation and propagation of elastic/acoustic waves can be safely ignored. Infinite rigidity is, of course, irreconcilable with the theory of relativity, as it produces elastic waves that propagate faster than the speed of light in vacuum.

The advantage of our approach is that it enables us to discuss the role of momentum entirely within the framework of the Maxwell-Lorentz theory of electrodynamics. We are thus able to prove the self-consistency of the theory as well as its consistency with the principles of conservation of energy, momentum, and angular momentum. Once the role of mechanical momentum within the theory of electrodynamics is clarified, it becomes only a matter of diligence to obtain, under less ideal but perhaps more practical circumstances, detailed solutions to the coupled equations of electrodynamics and elasticity, where one could give proper consideration to the energies associated with mechanical motions of the media, e.g., local as well as global translations, rotations and vibrations.

**2. Maxwell's macroscopic equations, the energy-momentum postulates, and the Lorentz law of force**. In the MKSA system of units, Maxwell's macroscopic equations are

$$\nabla \cdot \boldsymbol{D} = \rho_{\text{free}}, \tag{1a}$$

$$\nabla \times \boldsymbol{H} = \boldsymbol{J}_{\text{free}} + \partial \boldsymbol{D}/\partial t, \tag{1b}$$

$$\nabla \times \boldsymbol{E} = -\partial \boldsymbol{B}/\partial t, \tag{1c}$$

$$\nabla \cdot \boldsymbol{B} = 0. \tag{1d}$$

In these equations, electric displacement $\boldsymbol{D}$ and magnetic induction $\boldsymbol{B}$ are related to the polarization density $\boldsymbol{P}$ and magnetization density $\boldsymbol{M}$ via the identities

$$\boldsymbol{D} = \varepsilon_{\text{o}} \boldsymbol{E} + \boldsymbol{P}, \tag{2a}$$

$$\boldsymbol{B} = \mu_{\text{o}} \boldsymbol{H} + \boldsymbol{M}. \tag{2b}$$

In general, $\rho_{\text{free}}$, $\boldsymbol{J}_{\text{free}}$, $\boldsymbol{P}$, $\boldsymbol{M}$, $\boldsymbol{E}$, $\boldsymbol{H}$, $\boldsymbol{D}$, and $\boldsymbol{B}$ appearing in the above equations are functions of space and time $(\boldsymbol{r}, t)$ specified in an inertial frame of reference. As usual, $\rho_{\text{free}}$ and $\boldsymbol{J}_{\text{free}}$ are the densities of free charge and current, while $\mu_o = 4\pi \times 10^{-7}$ and $\varepsilon_o = 1/(c^2 \mu_o)$ are the permeability and permittivity of the free space [4-7,13,16,54].

It is important that one treat $\rho_{\text{free}}$, $\boldsymbol{J}_{\text{free}}$, $\boldsymbol{P}$ and $\boldsymbol{M}$ as sources of radiation, while considering $\boldsymbol{E}$ and $\boldsymbol{H}$ as the fields produced by these sources. There is no need to make any assumptions with regard to the nature of $\boldsymbol{P}$ and $\boldsymbol{M}$, whether these sources are locally or linearly related to the $\boldsymbol{E}$ and $\boldsymbol{H}$ fields, or even whether or not they are excited by the $\boldsymbol{E}$ and $\boldsymbol{H}$ fields. The material media are simply *defined* by $\rho_{\text{free}}$, $\boldsymbol{J}_{\text{free}}$, $\boldsymbol{P}$ and $\boldsymbol{M}$; it is the spatial and temporal variations of these sources that produce the $\boldsymbol{E}$ and $\boldsymbol{H}$ fields via Maxwell's equations.

At all locations, whether inside or outside the media, the rate of flow of electromagnetic energy (per unit area per unit time) is given by the Poynting vector $\boldsymbol{S}(\boldsymbol{r},t)$, which is defined as

$$\boldsymbol{S}(\boldsymbol{r},t) = \boldsymbol{E}(\boldsymbol{r},t) \times \boldsymbol{H}(\boldsymbol{r},t). \tag{3a}$$

Similarly, the densities of the electromagnetic momentum $\boldsymbol{p}_{\text{EM}}$ and angular momentum $\boldsymbol{L}_{\text{EM}}$ are universal functions of $\boldsymbol{E}$ and $\boldsymbol{H}$, which are given by the following formulas irrespective of the presence or absence of the sources $\rho_{\text{free}}$, $\boldsymbol{J}_{\text{free}}$, $\boldsymbol{P}$ and $\boldsymbol{M}$ at a given point $(\boldsymbol{r},t)$ in space and time:

$$\boldsymbol{p}_{\text{EM}}(\boldsymbol{r},t) = \boldsymbol{S}(\boldsymbol{r},t)/c^2, \tag{3b}$$

$$\boldsymbol{L}_{\text{EM}}(\boldsymbol{r},t) = \boldsymbol{r} \times \boldsymbol{p}_{\text{EM}}(\boldsymbol{r},t) = \boldsymbol{r} \times \boldsymbol{S}(\boldsymbol{r},t)/c^2. \tag{3c}$$

One cannot overemphasize the importance of the fact that the electromagnetic momentum densities given by Eqs.(3b) and (3c) are universal entities which, like the Poynting vector defined in Eq.(3a), must be considered an integral part of the classical theory of electrodynamics. One might even argue that, if their validity could not be ascertained within the Maxwell-Lorentz framework, then Eqs.(3) must be taken as fundamental postulates, on a par with Maxwell's equations themselves [51]. The strongest available argument in favor of the momentum density of Eq.(3b) is provided by a variant of the Einstein box "thought experiment" first proposed by Balazs [3], and reproduced here in Appendix A. Critics of the Balazs argument often cite its incompatibility with special relativity. These critics contend that the thought experiment implies that one end of the glass slab begins to move immediately after the arrival of the light pulse at the opposite end, irrespective of the length of the slab. We show, in the example presented in Sec.5, however, that this criticism is without merit. The slab in fact does *not* move as a solid block, rather it advances one atomic layer at a time. Meanwhile, accepting the Balazs argument does not automatically establish Eq.(3b) as a pillar of the electromagnetic theory, the reason being that Balazs's equating the mass of the light pulse with the pulse energy (normalized by $c^2$) lies outside the framework of the Maxwell-Lorentz theory. When a statement that is known to be true cannot be derived from the basic principles of a given theory, that statement itself should be considered a fundamental principle, hence our assertion that Eq.(3b) is a postulate – from which, incidentally, Eq.(3c) follows quite naturally.

It is worth mentioning here that the Abraham-Minkowski controversy persists precisely because the Maxwell-Lorentz theory is incapable of discriminating among several plausible expressions for the electromagnetic momentum. None of the arguments flowing from the various stress-energy tensor analyses, for instance, can assert the superiority of one form of momentum density over another; instead the usual conclusion of these analyses is that *all* the candidate expressions are valid provided that a corresponding force equation is used in conjunction with each momentum expression [26]. In contrast, we argue that Balazs's thought experiment leaves no room for ambiguity as to the nature of the electromagnetic momentum. However, since the classical Maxwell-Lorentz theory cannot be coerced into revealing Eq.(3b) as the true expression of momentum density, resort to postulation becomes inevitable.

It is reasonable to expect objections to accepting Eq.(3b) based solely on the evidence of a "thought experiment." The fact remains, however, that, upon fixing the linear and angular momentum densities at the values given by Eqs.(3b) and (3c), unique expressions will be obtained for the Lorentz force and torque densities, which are subsequently subject to experimental verification. Since laboratory experiments tend to measure force and torque rather than momentum and angular momentum, any experimental support for Eqs.(3b) and/or (3c) should come from careful comparisons of the measurement results against theoretical calculations of the electromagnetic force and torque experienced by material media.

Finally, one must specify the force and torque experienced by the material media in the presence of $\boldsymbol{E}$ and $\boldsymbol{H}$ fields at a given location. Since the media are *defined* by the sources $\rho_{\text{free}}$, $\boldsymbol{J}_{\text{free}}$, $\boldsymbol{P}$ and $\boldsymbol{M}$ contained within them, all one needs to know is the force and torque exerted by the fields on these sources. The generalized expressions of the Lorentz force and torque densities are [5,6,14,16,47-52]:

$$F(r,t) = \rho_{\text{free}} E + J_{\text{free}} \times \mu_o H + (P \cdot \nabla) E + (M \cdot \nabla) H + (\partial P/\partial t) \times \mu_o H - (\partial M/\partial t) \times \varepsilon_o E, \quad (4a)$$

$$T(r,t) = r \times F(r,t) + P(r,t) \times E(r,t) + M(r,t) \times H(r,t). \quad (4b)$$

Each term in the above expressions represents the force or torque exerted by either the *E*- or the *H*-field on individual sources $\rho_{\text{free}}$, $J_{\text{free}}$, *P* and *M*. Note, in particular, that each source is acted upon by a field, not by another source, even though different sources (e.g., $J_{\text{free}}$ and *M*, or *P* and *M*), might reside at the same location in space. We show in Appendix B that the conventional Lorentz force equation, $F=q(E+V\times B)$, can be rearranged to yield a similar, but by no means identical, expression to that given in Eq. (4a); the main difference between the two being a term, $\varepsilon_o \partial(E \times M)/\partial t$, whose absence from the conventional Lorentz law has been responsible for the so-called "hidden momentum" in electromagnetic systems containing a magnetization distribution $M(r,t)$ [8,9,14,47]. The modified force density of Eq. (4a), in conjunction with the momentum postulate of Eq. (3b), can be shown to be fully compatible with momentum conservation [52], whereas the traditional force law, $F=q(E+V\times B)$, leads to the "hidden momentum" absurdity.

In evaluating the force and torque exerted by the electromagnetic field on ponderable media, care must be taken that, in every instance, the relevant equations are solved self-consistently. For example, any motion imparted to the medium in consequence of the exertion of electromagnetic force and torque, which would result in a change of the spatio-temporal dependence of $\rho_{\text{free}}$, $J_{\text{free}}$, *P* and *M*, must be automatically incorporated into the solution of Maxwell's equations, solutions that relate the *E* and *H* fields to their sources $\rho_{\text{free}}$, $J_{\text{free}}$, *P* and *M*. Alternatively, if the electromagnetic fields are computed by assuming the sources $\rho_{\text{free}}(r,t)$, $J_{\text{free}}(r,t)$, $P(r,t)$ and $M(r,t)$, then the resulting force and torque on these sources, computed in accordance with Eqs. (4), cannot be allowed to further modify the sources; in particular, the exerted electromagnetic force and torque should not result in *additional* material motion, acoustic wave generation and propagation, etc., in a way that would modify the assumed strengths of the sources or their spatio-temporal dependences. This is *not* to say that mechanical motion and acoustic wave propagation should be ignored; rather it is stating the obvious that such motion must be treated self-consistently.

This completes our discussion of the fundamental tenets of the classical theory of electrodynamics, which is *all* that one needs for a discussion of energy and momentum in sufficiently rigid and sufficiently massive stationary media. Application of Eqs. (1–4) has yielded expressions for the electromagnetic and mechanical momenta of light pulses in diverse situations involving transparent, absorptive, dispersive, magnetic, and birefringent media in various combinations and configurations [18–23, 39–53]. The following sections provide a few examples of these solutions in situations which, despite the simplicity of the setup, reveal the essence of the momentum of light in its interaction with material objects.

**3. Reflection of light pulse from a conventional mirror**. Figure 1 shows a light pulse, propagating in the free space, in the process of reflection from a massive and solid flat mirror. The atomic layers at the surface of the mirror pick up momentum from the light pulse and transmit it to adjacent layers, which continue the process until all the layers are set in motion. The elastic wave, of course, propagates at the speed of sound within the mirror. Once the last atomic layer has been propelled forward and the vibrations died down, the motion of the mirror reaches steady state at a constant velocity *V*. Note that the mirror as a whole does not start to move once the light pulse arrives at its front facet; this would violate the principles of relativity by making the atomic layers at the rear move before the news of arrival of the light pulse at the front facet could have reached there.

The electromagnetic energy and momentum of the light pulse in the free space are $\mathcal{E}_{\text{pulse}}$ and $p_{\text{pulse}}=\mathcal{E}_{\text{pulse}}/c$. Denoting the mass of the mirror by *M*, its final momentum and kinetic energy will be $p_{\text{mirror}}=MV$ and $\mathcal{E}_{\text{mirror}}=\tfrac{1}{2}MV^2$. Assuming the mirror is massive enough, and expressing its kinetic energy as $\mathcal{E}_{\text{mirror}}=p_{\text{mirror}}^2/(2M)$, one can see immediately that, because of the large *M* in the denominator, the kinetic energy of the mirror can be neglected without requiring its momentum to vanish at the same time. Needless to say, if the mirror happens to be fastened to the ground, the mass of the Earth will become part of the mass of the mirror. The final momentum of the mirror will be $p_{\text{mirror}}=2p_{\text{pulse}}$, as required by momentum conservation. Also, since the energy transferred to the mirror is negligible, the reflected light pulse retains the entire energy of the incident pulse. Had the mirror been less massive, the reflected light would have been Doppler-shifted toward red, thus accounting for the kinetic energy of the mirror.

The transfer of momentum from the light pulse to the front facet of the mirror is mediated by Maxwell's macroscopic equations and the Lorentz law of force. If, for example, the mirror material is a high-electrical-conductivity metal such as gold or aluminum, the light penetrates the surface to within the skin-depth of the metal, typically a few nanometers, setting up a conduction current density $J_{\text{free}}$ in proportion to the local electric field $E(r,t)$. The magnetic field $H(r,t)$ of the light pulse, which also penetrates the surface to within the skin depth, then exerts a Lorentz force, $J_{\text{free}} \times \mu_o H$, on the surface current, propelling the atomic layers of the mirror forward [54].

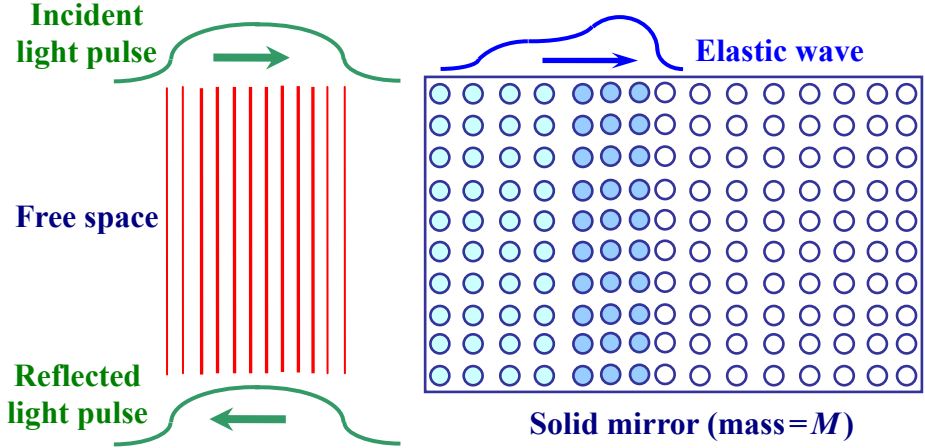

**Fig. 1**. A light pulse of energy $\mathcal{E}_{pulse}$ and momentum $\mathcal{E}_{pulse}/c$ is reflected from a solid mirror. The atomic layers at the surface pick up momentum from the pulse and transmit it to adjacent layers at the speed of sound within the mirror. Once the last atomic layer has been propelled forward and the vibrations died down, the motion of the mirror reaches steady state at a constant velocity $V$. The final momentum of the mirror is $MV = 2\mathcal{E}_{pulse}/c$.

**4. Light pulse entering a weakly absorbing medium**. Consider a finite-duration, finite-diameter light pulse in the free space, arriving at the surface of a weakly absorbing dielectric slab of refractive index $n$, as depicted in Fig. 2(a). Because of absorption, $n$ has a small imaginary component, which we shall ignore at first. For simplicity's sake, we also ignore the effects of dispersion, assuming that, within the spectral bandwidth of the pulse, the refractive index $n$ is frequency-independent.

The energy and momentum of the incident pulse are $\mathcal{E}^{(i)}_{pulse}$ and $\mathcal{E}^{(i)}_{pulse}/c$, respectively, and the Fresnel reflection coefficient at the front facet of the slab is $\rho = (1-n)/(1+n)$. With a sufficiently rigid and massive slab, the reflected pulse will not be Doppler shifted; its energy will be $\rho^2 \mathcal{E}^{(i)}_{pulse}$, and its momentum will be $-\rho^2 \mathcal{E}^{(i)}_{pulse}/c$, with the minus sign indicating that, upon reflection, the propagation direction is reversed. The optical energy transmitted into the slab, $\mathcal{E}^{(t)}_{pulse} = (1-\rho^2)\mathcal{E}^{(i)}_{pulse}$, will eventually be absorbed and converted to heat. With the absorption coefficient being small and the slab fairly thick, the transmitted pulse will reach a depth several times its own length before being fully extinguished.

The momentum difference between the incident and reflected light, once transferred to the slab, will remain in the slab ad infinitum. Upon reflection of the light pulse at its front facet, the total momentum of the slab, including, at first, the electromagnetic momentum of the transmitted pulse, will be $(1+\rho^2)\mathcal{E}^{(i)}_{pulse}/c$. In terms of the transmitted pulse energy, the total momentum acquired by the slab is thus $[(1+\rho^2)/(1-\rho^2)]\mathcal{E}^{(t)}_{pulse}/c = \frac{1}{2}(n+n^{-1})\mathcal{E}^{(t)}_{pulse}/c$.

Immediately after entering the slab, i.e., before being absorbed further down the road, the electromagnetic momentum of the light pulse will be $\mathcal{E}^{(t)}_{pulse}/(nc)$, as can be readily inferred from Eq. (3b) in conjunction with the fact that the speed of light inside the slab is $c/n$. This implies that the remaining momentum, $\frac{1}{2}(n-n^{-1})\mathcal{E}^{(t)}_{pulse}/c$, must somehow have entered the slab as mechanical momentum. Frame (b) of Fig. 2 shows the mechanism of this momentum entry into the slab. As the leading edge of the pulse enters the medium through the front facet, it exerts a Lorentz force on the oscillating atomic/molecular dipoles of the material, which oscillations are induced by the leading edge itself [39,40,46]; the term responsible for this Lorentz force in Eq. (4a) is $(\partial \mathbf{P}/\partial t) \times \mu_o \mathbf{H}$. The orange-colored molecules in Fig. (2b) thus acquire a mechanical momentum and begin to move forward. By the time the trailing edge of the pulse enters the slab, all the molecules that have already seen the leading edge are moving forward.

The trailing edge has the opposite effect on these molecules: it exerts a braking Lorentz force in the reverse direction that slows the molecules down; see frame (c) of Fig. 2. The term in Eq. (4a) that is responsible for the decelerating force of the trailing edge is the same one that gives rise to the accelerating force of the leading edge, namely, $(\partial \mathbf{P}/\partial t) \times \mu_o \mathbf{H}$. At this point, assuming the light pulse is sufficiently short, the elastic forces have not yet had time to exert much influence, so the orange-colored molecules, which are moving forward, carry most of the mechanical momentum. Behind them are the dark-blue-colored molecules, which have been displaced forward by the light pulse that has just passed through them. Because the medium is weakly absorbing, the dark-blue-colored molecules retain a small fraction of their initial momentum. These dark-blue-colored molecules now begin to drag along, via elastic forces, the adjacent white molecules that have been left behind. At the same time, the dark-blue-colored molecules are pulled backward by the elastic reaction force. We mention in passing that, had there been no absorption of light by the medium, the forces exerted by the leading and trailing edges of the pulse would have been exactly equal and opposite, leaving the dark-blue-colored molecules displaced but without momentum.

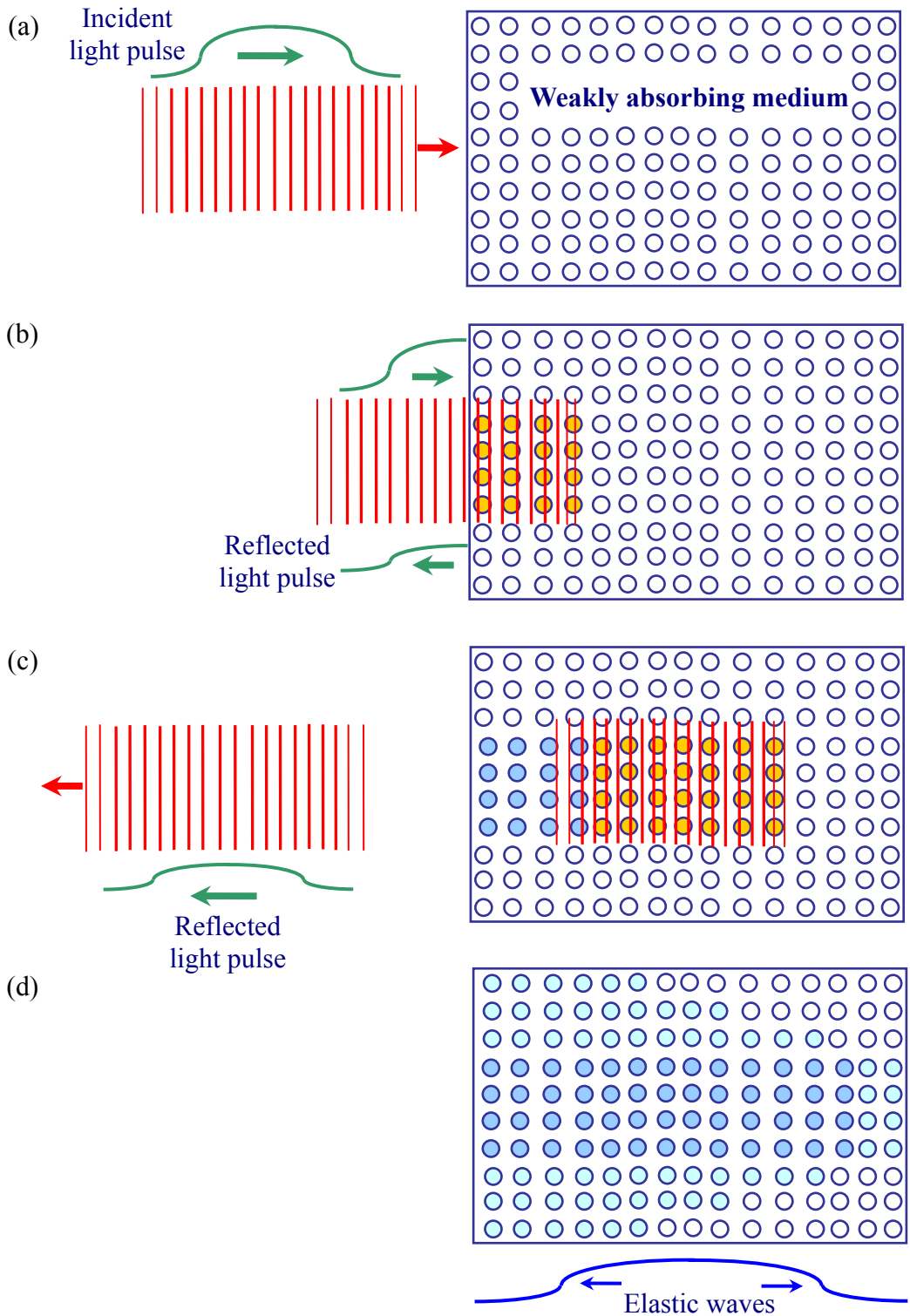

**Fig. 2**. (a) A light pulse of energy $\mathcal{E}^{(i)}_{pulse}$ and momentum $\mathcal{E}^{(i)}_{pulse}/c$ arrives at the front facet of a weakly-absorbing dielectric slab of refractive index $n$. (b) As the leading edge of the pulse enters the slab, it exerts a Lorentz force on the atomic dipoles; the orange-colored molecules thus pick up momentum and move forward. The trailing edge will have the opposite effect on these molecules: it exerts a braking Lorentz force that slows them down. (c) The dark-blue-colored molecules, having been displaced forward by the light pulse while still retaining some momentum, try to drag along the adjacent white molecules via elastic forces; at the same time, these dark-blue-colored molecules are further slowed down and pulled back by the elastic reaction force of their neighbors. (d) Later on, when the light has been extinguished within the medium, the dark-blue-colored molecules, which have been displaced forward by the light pulse, manage to drag along some of the remaining molecules (light-blue-colored). The displaced molecules also push forward those molecules that lie ahead of them. Proceeding at the speed of sound, this collective process eventually comes to equilibrium. The end sees the entire slab moving forward with a uniform velocity $V$, which, because of the large mass $M$ of the slab, is very small. While the acquired mechanical momentum $MV$ of the slab is significant, its kinetic energy $\frac{1}{2}MV^2$ is negligible.

Frame (d) of Fig. 2 shows the situation at a later time, when the light pulse within the slab has been fully extinguished. The dark-blue-colored molecules, which have been displaced forward by the light pulse, have pulled forward some of the remaining molecules (light-blue-colored); they have also begun to push forward the molecules that lie in front of them. Thus the elastic forces, acting between adjacent molecules, distribute the molecular motion and displacement of the dark-blue-colored molecules among the remaining molecules of the slab. This process, which proceeds at the speed of sound within the medium, eventually comes to an equilibrium, with the slab settling into a uniform forward motion with a mechanical momentum $MV = (1+\rho^2)\mathcal{E}_{\text{pulse}}^{(i)}/c$.

Note that in the entire process described above, there are no special forces acting on the front facet of the slab; aside from the slow attenuation of the light pulse as it propagates forward, all molecular layers are subject to the same Lorentz force. The light pulse deposits its momentum, which initiates molecular motion and displacement, at the supersonic speed of $c/n$. The acoustic waves, produced and propagated by the elastic forces that act between adjacent molecules, come into play immediately afterward, helping to redistribute the mechanical momentum and molecular displacements produced in the wake of the light pulse, without introducing any new momentum of their own. The equality of action and reaction makes the net contribution to momentum of the elastic forces precisely equal to zero. The only source of mechanical momentum is thus seen to be the Lorentz force of the electromagnetic field, acting on bound currents, i.e., oscillating dipoles of the dielectric medium. In the final analysis, it is the momentum of the electromagnetic field that, through the action of the Lorentz force, is converted to mechanical momentum of the material medium [39, 51].

**5. Passage of a light pulse through a transparent slab**. A light pulse having energy $\mathcal{E}_{\text{pulse}}$ and momentum $\mathcal{E}_{\text{pulse}}/c$ goes through a transparent slab of refractive index $n$, as shown in Fig. 3. The front and rear facets of the slab are anti-reflection (AR) coated, so that the entire pulse can enter the slab from the free space on the left, then re-emerge into the free space on the right, without losing any of its energy or momentum. Since the material of the slab is transparent, its refractive index $n$ must be real-valued. To simplify the argument, we shall also assume that the material is free from dispersion, although, strictly speaking, this assumption is not necessary.

During the entry phase, interference fringes are set up within the AR-layer, producing a net Lorentz force in the direction pointing away from the slab. This force, when integrated over the pulse duration, gives the slab a net backward mechanical momentum of $-\frac{1}{2}(n-1)^2\mathcal{E}_{\text{pulse}}/(nc)$ [39]; the short arrows in Fig. 3(b) indicate the momentum acquired by the AR-layer, which will eventually be distributed throughout the entire slab. At the same time, the leading edge of the light pulse presses against the molecules of the slab, giving them a net forward momentum [39].

The orange-colored molecules in Figs. 3(a, b) carry the mechanical momentum of the light pulse inside the slab. Once the pulse is fully inside, the force acting on the AR-layer vanishes; the AR-layer retains its acquired mechanical momentum at first, then begins to share it with the rest of the slab through the action of elastic forces. Within the slab, the trailing edge of the pulse brings the molecules to a complete stop, as indicated by the blue-colored molecules in Fig. 3(b). At this time, the mechanical momentum of the orange-colored molecules is $\frac{1}{2}(n-n^{-1})\mathcal{E}_{\text{pulse}}/c$ [39, 46, 49], which, together with the electromagnetic, i.e., Abraham, momentum of the light pulse, amounts to a total pulse momentum of $\frac{1}{2}(n+n^{-1})\mathcal{E}_{\text{pulse}}/c$. Adding to this the mechanical momentum of the AR-layer brings the overall momentum of the system to $\mathcal{E}_{\text{pulse}}/c$, which is precisely the electromagnetic momentum the pulse had prior to entering the slab. The total momentum is thus seen to be conserved.

The light pulse eventually reaches the exit facet and emerges into the free space on the right-hand side. The interference fringes that are set up within the AR-layer at the exit fact give a net forward momentum to this AR-layer, as indicated by the short arrows in Fig. 3(c). This momentum is equal in magnitude and opposite in direction to that imparted to the AR-layer at the entrance facet [39, 47]. With the departure of the light pulse, the net mechanical momentum of the slab returns to zero, although all the molecules that have been touched by the light, i.e., blue-colored molecules, are now displaced forward. Elastic forces now enter the picture and readjust the molecular positions until the internal forces are once again balanced. No new momentum, of course, can be generated inside the slab as a result of these elastic forces pushing and pulling on adjacent molecules.

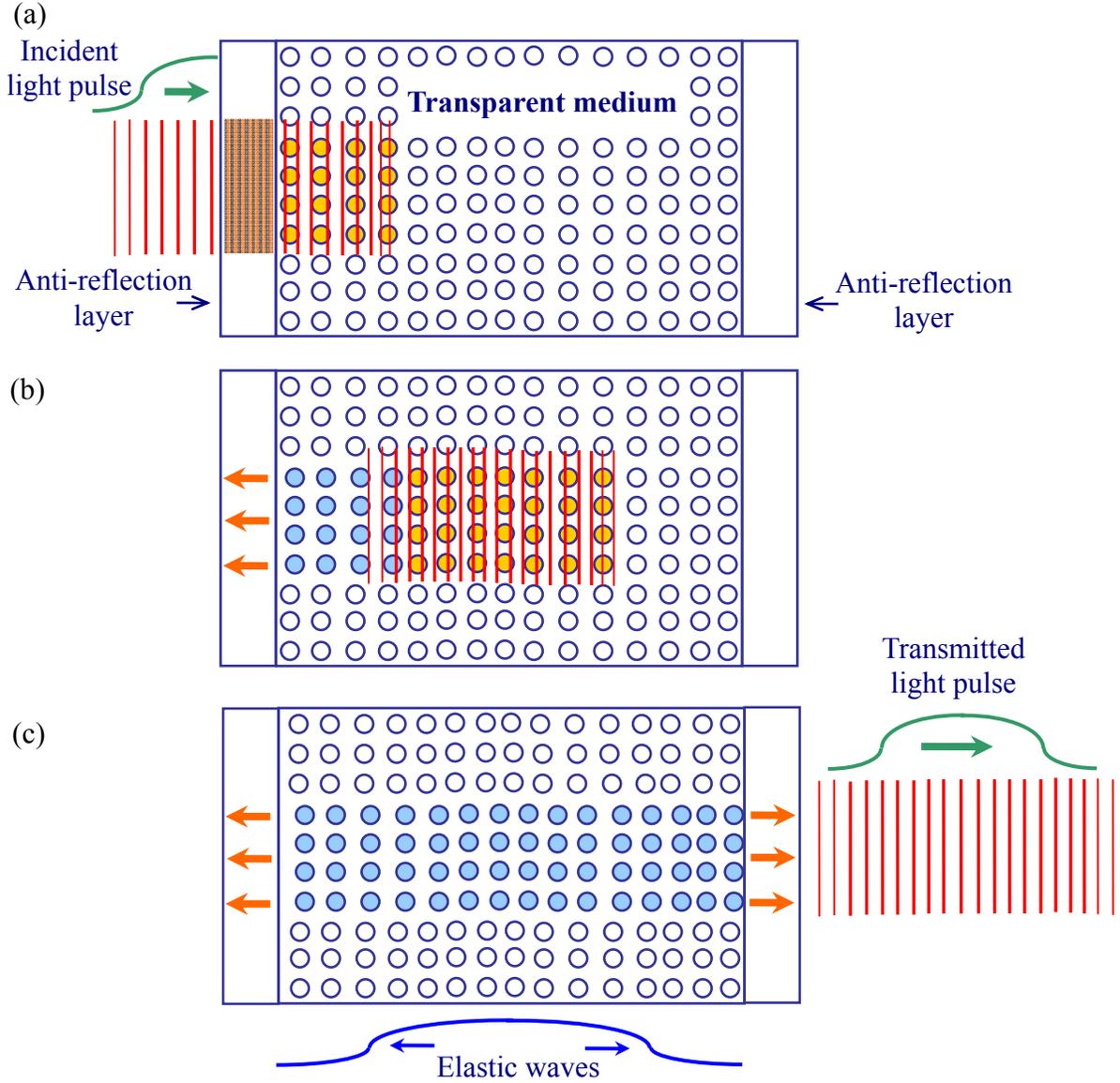

**Fig. 3**. A light pulse of energy $\mathcal{E}_{\text{pulse}}$ and momentum $\mathcal{E}_{\text{pulse}}/c$ goes through a transparent slab of refractive index $n$. Both ends of the slab are AR-coated. When the pulse enters the slab, and also when it exits, interference fringes are set up within the AR-layers that exert a net Lorentz force on each such layer; the short arrows in (b) and (c) indicate the equal but opposite mechanical momenta thus imparted to the AR-layers. Propelled by the leading edge of the light pulse, the orange-colored molecules of the slab in (a) and (b) carry the mechanical momentum within the slab. In due time, the trailing edge of the pulse brings these molecular motions to a halt, but not before the molecules have been slightly displaced forward; in (b) and (c) the displaced molecules are shown as blue-colored. Immediately after the pulse exits through the rear facet in (c), the slab molecules that have been touched by the light, i.e., blue-colored molecules, are seen to have been displaced forward. Subsequently, elastic forces take over and adjust the molecular positions until all internal forces are balanced.

With the pulse out of the slab, the energy and momentum are once again entirely within the light pulse. The slab, after internal adjustments, comes to a halt, albeit with its center of mass slightly displaced to the right. The larger the mass $M$ of the slab, the smaller will be its forward displacement $\Delta$, although, all else being equal, the product $M\Delta$ will be a constant. To determine the value of this constant, note that the center of mass of the system (i.e., light pulse + slab) along the propagation direction must follow the same path as it would have, had the slab been shifted vertically (or sideways) out of the pulse's path from the beginning. When inside the slab, the pulse slows down from its vacuum speed of $c$ to $c/n$. Denoting the length of the slab by $L$, the pulse spends a total time of $nL/c$ inside. During this time, the light pulse falls behind by a distance of $c(1-n^{-1})(nL/c)$ relative to an identical light pulse traveling outside the slab. The product of this distance and the mass $\mathcal{E}_{\text{pulse}}/c^2$ of the light pulse, namely, $(n-1)L\mathcal{E}_{\text{pulse}}/c^2$, must be equal to $M\Delta$, the product of mass and displacement for the slab, if the center of mass of the system is to have the expected behavior. The displacement of the slab is thus found to be $\Delta=(n-1)L\mathscr{E}_{\text{pulse}}/(Mc^2)$.

The above argument, which is a variant of the Einstein box gedanken experiment [54], was first proposed by Balazs in 1953 [3]. A major conclusion reached through this type of thought experiment is that, inside a homogeneous transparent medium, the electromagnetic component of the momentum of light must be $\mathscr{E}_{\text{pulse}}/(n_g c)$, where $n_g$ is the group refractive index of the material medium. This conclusion is in perfect agreement with Eq.(3b), as may be seen by denoting the cross-sectional area of the light pulse by $A$, its duration by $\tau$, and the magnitude of its Poynting vector by $S$. The energy of the pulse will then be $\mathscr{E}_{\text{pulse}}=SA\tau$, its length $(c/n_g)\tau$, its volume $A(c/n_g)\tau$, and its electromagnetic momentum, in accordance with Eq.(3b), $A(c/n_g)\tau S/c^2 = \mathscr{E}_{\text{pulse}}/(n_g c)$.

**6. Concluding remarks.** While the momentum of light inside material media is ultimately tied to the inertial and elastic properties of the media, we have shown that, in circumstances where the media are sufficiently rigid and massive, the momentum of a light pulse can be described in terms solely of the electromagnetic properties of the media. The question then becomes one of the consistency of the equations of classical electrodynamics, not only among themselves, but also with the conservation laws of energy, momentum, and angular momentum. This question has now been answered in the affirmative, provided that Eqs.(1−4) of the macroscopic Maxwell-Lorentz theory are applied consistently across the board, within the volumes of the material media as well as at their surfaces and interfaces. Any changes in the total linear (or angular) electromagnetic momentum of a closed system, obtained by integrating the momentum density of Eq.(3b) (or Eq.(3c)) over all space, will result in a corresponding mechanical force (or torque) in accordance with the Lorentz law of Eq.(4a) (or Eq.4(b)). This back and forth exchange between electromagnetic momentum and mechanical force can be shown to conserve the total linear and angular momenta of the system, irrespective of the specific properties of the media, i.e., whether or not the media are homogeneous, absorptive, dispersive, isotropic, linear, magnetic, etc.

Historically, a strong argument in favor of the Minkowski momentum has been based on the measured values of radiation pressure on a high-reflectivity, flat mirror submerged in a liquid dielectric [28,33]. In fact, within the framework of the Maxwell-Lorentz theory, one finds that a light pulse propagating in a transparent, dispersionless liquid of refractive index $n$, when reflected from a submerged mirror, imparts a total momentum of $2n\mathscr{E}_{\text{pulse}}/c$ to the mirror. In other words, inside the host liquid, the light "appears" to carry the Minkowski momentum. In reality, however, the difference between the "Minkowski momentum" transmitted to the mirror and the actual momentum of the pulse is transferred to the liquid host via Lorentz forces that are set up in the region of overlap between the incident and reflected light pulses [39,45,53]. To the best of our knowledge, the role of the medium in momentum transfer was first proposed by R. V. Jones, although no specific mechanisms were suggested [55]. In addition to our own analysis of the photon drag effect [40], of the submerged mirror experiments [53], of the optical trapping of micro-beads under focused laser light [56], and of the radiation force on a liquid surface [44], a recent theoretical paper by Pfeifer *et al* describes the conditions under which the Minkowski momentum could "appear" to be responsible for the experimental observations [27]. We have also shown, through a detailed analysis of the observed force at the end-face of a nano-filament [57-59], how the combination of electromagnetic and mechanical forces could give rise to observed phenomena that do not lend themselves to a simple interpretation in terms of either the Abraham or the Minkowski momentum. It is perhaps safe to say at this point that the Minkowski momentum does not seem to have an independent existence, but a combination of factors could conspire to give the "appearance" of the Minkowski momentum in certain experimental settings. We feel confident that a thorough analysis of the experimental results of Walker *et al* [32], Kristensen and Woerdman [36], and Campbell *et al* [37] would similarly identify the role of momentum transfer to the medium as responsible for the appearance of the Minkowski momentum in these experiments.

The results of this paper can be readily extended to cover magnetic media specified by their permittivity $\varepsilon_o \varepsilon(\omega)$ and permeability $\mu_o \mu(\omega)$ [47−50]. For a light pulse entering a magnetic slab at normal incidence, the Fresnel reflection coefficient at the entrance facet will be $\rho=(1-\eta)/(1+\eta)$, where $\eta=\sqrt{\varepsilon/\mu}$ is the admittance of the medium. Conservation of momentum then shows, following essentially the same argument as presented in Section 4, that the total momentum of the pulse inside the slab will be $p_{\text{total}}=\tfrac{1}{2}(\eta+\eta^{-1})\mathscr{E}^{(t)}_{\text{pulse}}/c$, where $\mathscr{E}^{(t)}_{\text{pulse}}$ is the energy of the pulse transmitted into the slab [47]. Note, in particular, that for a negative-index medium, where both $\varepsilon$ and $\mu$ are real-valued and negative, their ratio will be positive, resulting in a real-valued and positive $\eta$. As before, in compliance with the Balazs argument [3], the electromagnetic momentum of this pulse turns out to be $p_{\text{EM}}=\mathscr{E}^{(t)}_{\text{pulse}}/(n_g c)$, where the group refractive index $n_g$ is related to the phase refractive index $n(\omega)=\sqrt{\varepsilon\mu}$ through the well-known relation $n_g=d[\omega n(\omega)]/d\omega$. Once again, we find $p_{\text{mech}}=p_{\text{total}}-p_{\text{EM}}$ [47].

## Appendix A

A light pulse traveling inside a transparent medium has both electromagnetic and mechanical momenta. The mechanical momentum is due to the motion of the atoms/molecules of the medium in response to the electromagnetic forces exerted upon them by the light pulse. In a variant of the "Einstein box" Gedanken experiment first proposed by Balazs [3], a pulse of energy $E = mc^2$ and (free-space) momentum $\mathbf{p} = mc\hat{\mathbf{z}}$ travels either outside or inside a transparent slab of length $L$ and mass $M_o$; see Fig. A1. The entrance and exit facets of the slab are anti-reflection coated to ensure the passage of the entire pulse through the slab, with no reflection losses whatsoever. The pulse crosses the slab in a time interval $\Delta t = L/V_g$, where the group velocity $V_g$ is a function of the optical frequency $f$ and the (frequency-dependent) material parameters $\varepsilon$ and $\mu$.

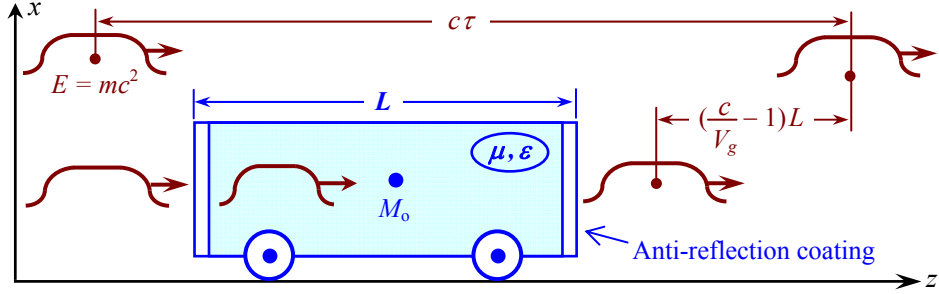

**Fig. A1**. The Balazs thought experiment featuring a short pulse of light and a transparent slab of length $L$ and mass $M_o$. In the free-space region outside the slab, the pulse, having energy $E = mc^2$ and momentum $\mathbf{p} = mc\hat{\mathbf{z}}$, travels with speed $c$. Inside the slab, the pulse travels with the group velocity $V_g$. The entrance and exit facets of the slab are anti-reflection coated to ensure the passage of the entire pulse through the slab. In one experiment, the pulse travels entirely in the free-space region outside the slab, while in another, the pulse spends a fraction of its time inside the slab. Since no external forces are at work, the center of mass of the system (consisting of the light pulse and the slab) must be displaced equally in the two experiments.

When the pulse travels outside and the slab is stationary, the center of mass of the system moves along the $z$-axis at the constant velocity $V_{CM} = mc/(m + M_o)$. The displacement of the center of mass during a time interval $\tau$ is, therefore, $mc\tau/(m + M_o)$. If the pulse goes through the slab, however, its velocity, while inside the slab, will drop down to the group velocity $V_g$. The emergent pulse will thus stay behind the pulse that has traveled in the free-space by a distance $[(c/V_g) – 1]L$, as shown in Fig. A1. Therefore, for the system's center of mass to be in the same place in both experiments, it is necessary for the slab in the latter case to have shifted to the right by $\Delta z = [(c/V_g) – 1]Lm/M_o$. This displacement, which occurs during the time interval $\Delta t = L/V_g$ when the pulse is inside the slab, requires the slab's mechanical momentum during the passage of the pulse to be

$$\mathbf{p}_M = M_o(\Delta z/\Delta t)\hat{\mathbf{z}} = m(c - V_g)\hat{\mathbf{z}}. \tag{A1}$$

Since the total momentum of the system is that of the pulse before entering the slab, namely, $\mathbf{p} = mc\hat{\mathbf{z}}$, we conclude that the pulse's electromagnetic momentum inside the slab must be

$$\mathbf{p}_E = mV_g\hat{\mathbf{z}} = (E/c)(V_g/c)\hat{\mathbf{z}}. \tag{A2}$$

In other words, the pulse's electromagnetic momentum within the medium is reduced by a factor of $V_g/c$ relative to its free-space value. In a dispersionless medium where $V_g = c/n$, $n$ being the refractive index, the electromagnetic momentum of the pulse will be $\mathbf{p}_E = (E/nc)\hat{\mathbf{z}}$ – commonly referred to as the Abraham momentum. The difference between the free-space momentum of the pulse and its electromagnetic (or Abraham) momentum is thus transferred to the slab in the form of mechanical momentum, $\mathbf{p}_M$, causing the slab's eventual displacement in a manner consistent with the demands of the Einstein box experiment.

## Appendix B

The Lorentz law of force, $\mathbf{f} = q(\mathbf{E} + \mathbf{V} \times \mathbf{B})$, which specifies the force exerted by the electromagnetic fields $(\mathbf{E}, \mathbf{B})$ on a particle of charge $q$ and velocity $\mathbf{V}$, may be written as an expression of the force density $\mathbf{F}(\mathbf{r}, t)$ exerted at $(\mathbf{r}, t)$ on the local charge and current densities, $\rho_{\text{total}}(\mathbf{r}, t)$, $\mathbf{J}_{\text{total}}(\mathbf{r}, t)$, as follows:

$$\mathbf{F}(\mathbf{r}, t) = \rho_{\text{total}}(\mathbf{r}, t)\mathbf{E}(\mathbf{r}, t) + \mathbf{J}_{\text{total}}(\mathbf{r}, t) \times \mathbf{B}(\mathbf{r}, t). \tag{B1}$$

Here $\rho_{total}(r,t) = \rho_{free}(r,t) - \nabla \cdot P(r,t)$ incorporates the densities of free and bound charges. Similarly, $J_{total}(r,t) = J_{free}(r,t) + \partial P(r,t)/\partial t + \mu_o^{-1} \nabla \times M(r,t)$ contains the densities of free as well as bound currents arising from polarization and magnetization densities. The Lorentz force density of Eq.(B1) may thus be written

$$F(r,t) = \rho_{free}E - (\nabla \cdot P)E + J_{free} \times \mu_o H + (\partial P/\partial t) \times \mu_o H + (\nabla \times M) \times H + J_{total} \times M. \tag{B2}$$

The fact that Maxwell's Eq.(1b) may be written equivalently as $\nabla \times B = \mu_o J_{total} + \mu_o \varepsilon_o \partial E/\partial t$, allows one to rewrite Eq.(B2) as follows:

$$F = \rho_{free}E + J_{free} \times \mu_o H - (\nabla \cdot P)E + (\partial P/\partial t) \times \mu_o H + (\nabla \times M) \times H + (\nabla \times H) \times M + \mu_o^{-1}(\nabla \times M) \times M - \varepsilon_o(\partial E/\partial t) \times M. \tag{B3}$$

Then, with the help of the vector identity $\nabla(A \cdot B) = (A \cdot \nabla)B + (B \cdot \nabla)A + A \times (\nabla \times B) + B \times (\nabla \times A)$, we will have

$$F(r,t) = \rho_{free}E + J_{free} \times \mu_o H - (\nabla \cdot P)E + (\partial P/\partial t) \times \mu_o H + (M \cdot \nabla)H + (H \cdot \nabla)M - \nabla(M \cdot H)$$
$$+ \mu_o^{-1}(M \cdot \nabla)M - \tfrac{1}{2}\mu_o^{-1}\nabla(M \cdot M) - (\partial M/\partial t) \times \varepsilon_o E - \varepsilon_o \partial(E \times M)/\partial t. \tag{B4}$$

Next, we combine and rearrange the terms and, noting from Maxwell's Eq.(1d) that $\nabla \cdot B = 0$, we add the null term $\mu_o^{-1}(\nabla \cdot B)M$ to the right-hand side of Eq.(B4) to obtain

$$F(r,t) = \rho_{free}E + J_{free} \times \mu_o H + (P \cdot \nabla)E + (\partial P/\partial t) \times \mu_o H + (M \cdot \nabla)H - (\partial M/\partial t) \times \varepsilon_o E - \varepsilon_o \partial(E \times M)/\partial t$$
$$- [(P \cdot \nabla)E + (\nabla \cdot P)E] + \mu_o^{-1}[(B \cdot \nabla)M + (\nabla \cdot B)M] - \mu_o^{-1} \nabla[M \cdot (B - \tfrac{1}{2}M)]. \tag{B5}$$

The last three terms of the above equation are complete differentials, that is,

$$(P \cdot \nabla)E + (\nabla \cdot P)E = \partial(P_x E)/\partial x + \partial(P_y E)/\partial y + \partial(P_z E)/\partial z, \tag{B6a}$$

$$(B \cdot \nabla)M + (\nabla \cdot B)M = \partial(B_x M)/\partial x + \partial(B_y M)/\partial y + \partial(B_z M)/\partial z, \tag{B6b}$$

$$\nabla[M \cdot (B - \tfrac{1}{2}M)] = [(\partial/\partial x)\hat{x} + (\partial/\partial y)\hat{y} + (\partial/\partial z)\hat{z}][M \cdot (B - \tfrac{1}{2}M)]. \tag{B6c}$$

This means that, if $F(r,t)$ of Eq.(B5) is integrated over the entire volume of space that contains material media, taking note of the fact that both $P$ and $M$ vanish outside the media, one finds that the net contribution to *total* force of the last three terms amounts to zero. Thus, as far as the *total* force exerted on the material media of a given electromagnetic system is concerned, the last three terms of Eq.(B5) are inconsequential and may as well be dropped. We are then left with

$$F(r,t) = \rho_{free}E + J_{free} \times \mu_o H + (P \cdot \nabla)E + (\partial P/\partial t) \times \mu_o H + (M \cdot \nabla)H - (\partial M/\partial t) \times \varepsilon_o E - \varepsilon_o \partial(E \times M)/\partial t. \tag{B7}$$

The last term in the above equation can be shown to be responsible for the so-called "hidden" momentum [47]. To ensure momentum conservation in classical electrodynamics, it is thus necessary to remove from Eq.(B7) the term $-\varepsilon_o \partial(E \times M)/\partial t$. After all, the Lorentz law being only a *postulate* of the classical theory, its incompatibility with momentum conservation provides sufficient grounds for demanding its modification. Removing the last term from Eq.(B7) leaves behind a modified form of the Lorentz law that has a simple interpretation in terms of the localized and instantaneous forces exerted by the fields $E(r,t)$ and $H(r,t)$ on the material media *defined* by $\rho_{free}(r,t)$, $J_{free}(r,t)$, $P(r,t)$ and $M(r,t)$. This modification can be shown to be compatible with Maxwell's equations, with the conservation laws of energy, momentum, and angular momentum, and with the special theory of relativity [52].

The fact that the last term in Eq.(B7), as well as the last three terms in Eq.(B5), must be dropped from the conventional expression of the Lorentz law indicates that $P$ and $M$ are *not* reducible to bound-charge and bound-current densities (as implied by Maxwell's macroscopic equations). We conclude that $P$ and $M$ must be treated as independent sources of the electromagnetic field. The modified expressions of the Lorentz force and torque densities given by Eqs.(4) in conjunction with the energy flux density $S(r,t)$, the linear momentum density $p_{EM}(r,t)$, and the angular momentum density $L_{EM}(r,t)$ given by Eqs.(3), thus complete the foundational postulates of the classical theory. Maxwell's macroscopic equations, Eqs.(1) and (2), in conjunction with the energy and momentum postulates of Eqs.(3) and the force and torque expressions of Eqs.(4) provide a complete and consistent basis for treating *all* electromagnetic phenomena in the classical domain.

**Acknowledgements.** This work has been supported by the Air Force Office of Scientific Research (AFOSR) under contract number FA 9550-04-1-0213. The author is grateful to Armis Zakharian and John Weiner for many helpful discussions. He also thanks the anonymous referees whose substantive suggestions and extensive comments on several aspects of the manuscript have significantly improved the paper.